\documentclass[traditabstract]{aa}
\usepackage{txfonts}
\usepackage{graphicx}
\usepackage{natbib}
\bibpunct{(}{)} {;}{a}{ }{,}
\begin{document}

\title{Near-infrared spectroscopy of a nitrogen-loud quasar\\ 
       SDSS J1707+6443}

\author{N. Araki\inst{1}
          \and
          T. Nagao\inst{2,3}
          \and
          K. Matsuoka\inst{1,3}
          \and
          A. Marconi\inst{4}
          \and
          R. Maiolino\inst{5,6}
          \and
          H. Ikeda\inst{1,3} 
          \and \\
          T. Hashimoto\inst{7}
          \and
          Y. Taniguchi\inst{8}
          \and
          T. Murayama\inst{9}
          }
\institute{
   Graduate School of Science and Engineering, Ehime University, 
   2-5 Bunkyo-cho, Matsuyama 790-8577, Japan
   \and
   The Hakubi Center for Advanced Research, Kyoto University, 
   Yoshida-Ushinomiya-cho, Sakyo-ku, Kyoto 606-8302, Japan
   \and
   Department of Astronomy, Graduate School of Science, Kyoto University, 
   Kitashirakawa-Oiwake-cho, Sakyo-ku, Kyoto 606-8502, Japan
   \and
   Dipartimento di Fisica e Astronomia, Universit\`a degli Studi di Firenze, 
   Largo E. Fermi 2, 50125, Firenze, Italy
   \and
   INAF -- Osservatorio Astrofisico di Roma, 
   Via di Frascati 33, 00040 Monte Porzio Catone, Italy
   \and
   Cavendish Laboratory, Univerisy of Cambridge, 19 J. J. Thomson Ave., 
   Cambridge CB3 0HE, UK
   \and
   Optical and Infrared Astronomy Division, National Astronomical 
   Observatory of Japan, 2-21-1 Osawa, Mitaka 181-8588, Japan
   \and
   Research Center for Space and Cosmic Evolution, Ehime University, 
   2-5 Bunkyo-cho, Matsuyama 790-8577, Japan
   \and
   Astronomical Institute, Graduate School of Science, Tohoku University, 
   Aramaki, Aoba, Sendai 980-8578, Japan
   }
\date{Received xxx; accepted xxx}

\abstract{
We present near-infrared spectroscopy of the $z\simeq 3.2$ quasar SDSS 
J1707+6443, obtained with MOIRCS on the Subaru Telescope.
This quasar is classified as a ``nitrogen-loud'' quasar because of
the fairly strong N~{\sc iii}] and N~{\sc iv}] semi-forbidden emission
lines from the broad-line region (BLR) observed in its rest-frame UV spectrum.
However, our rest-frame optical spectrum from MOIRCS shows strong 
[O~{\sc iii}] emission from the narrow-line region (NLR) suggesting that, at 
variance with the BLR, NLR gas is not metal-rich.
In order to reconcile these contradictory results, there may be two
alternative possibilities;
(1) the strong nitrogen lines from the BLR are simply due to a very 
high relative abundance of nitrogen rather than to a very high BLR metallicity, 
or (2) the BLR metallicity is not representative of the metallicity of the host   
galaxy, better traced by the NLR. In either case, the strong broad nitrogen lines 
in the UV spectrum are not indication of a
chemically enriched host galaxy. 
We estimated the black hole mass and Eddington ratio of
this quasar from the velocity width of both C~{\sc iv} and H$\beta$, that
results in log($M_{\rm BH}$$/$$M_{\odot}$) = 9.50 and 
log($L_{\rm bol}$$/$$L_{\rm Edd}$) = -0.34.
The relatively high Eddington ratio is consistent with our earlier result that 
strong nitrogen emission from BLRs is associated with high Eddington ratios. 
Finally, we detected significant [Ne~{\sc iii}] emission from the NLR, implying 
a quite high gas density of 
$n_{\rm e} \sim 10^6$ cm$^{-3}$ and suggesting a strong coupling between 
quasar activity and dense interstellar clouds in the host galaxy.
}
\keywords{
   Galaxies: active - 
   Galaxies: nuclei - 
   quasars: emission lines - 
   quasars: individual: SDSS J1707+6443
}
\maketitle

\section{Introduction}

The chemical composition of galaxies is a powerful tool to distinguish various 
evolutionary scenarios as it is the result of past star-formation activity and 
is also affected by gas inflows and outflows. The chemical properties of active 
galactic nuclei (AGN) are particularly interesting, because the huge luminosities 
of AGN enable us to accurately measure spectroscopic features of quasars 
even at high redshifts and thus to explore the chemical properties in the early 
universe. Another interesting aspect of assessing the chemical composition
of AGNs is that it provides clues on the co-evolution between 
galaxies and their supermassive black holes (SMBHs), that has 
been inferred by the tight correlation between the mass of SMBHs
(M$_{\rm BH}$) and their host spheroidal observed in the local Universe
\citep[e.g.,][]{2003ApJ...589L..21M,2000ApJ...539L...9F,2000ApJ...539L..13G}. 
\citet{2011A&A...527A.100M} reported a tight relationship between the 
metallicity in broad-line regions (BLRs) and $M_{\rm BH}$ at $z \sim 2 - 3$, 
suggesting that the evolution of SMBHs is associated with the cumulative star 
formation in the host galaxies \citep[see also][]{2004ApJ...608..136W,
2006A&A...447..157N}.

A widely adopted method to estimate the BLR metallicity ($Z_{\rm BLR}$) 
exploits the flux of nitrogen emission lines such as N~{\sc v}$\lambda$1240. 
Nitrogen is a secondary element and thus its relative abundance is proportional 
to the metallicity, [N/O] $\propto$ [O/H] or, equivalently, [N/H] $\propto$ 
[O/H]$^2$ \citep[see, e.g.,][]{1992ApJ...391L..53H}. Based on this method, it has 
been found that the BLR in most quasars show super-solar metallicities 
\citep[e.g.,][]{2006A&A...447..157N}, even for quasars at $z \gtrsim 6$ 
\citep{2007AJ....134.1150J,2009A&A...494L..25J,2011Natur.474..616M}. The 
BLR metallicity is up to $Z_{\rm BLR} \sim 10 Z_\odot$ in the most extreme 
cases although it seems hard to understand such high metallicities through 
ordinary galaxy chemical evolution models \citep[e.g.,][]{1993ApJ...418...11H}.
More interestingly, some quasars show extremely strong emission in N~{\sc v} 
and in other nitrogen lines (especially N~{\sc iv}]$\lambda$1486 and 
N~{\sc iii}]$\lambda$1750 semi-forbidden lines). The estimated $Z_{\rm BLR}$ 
reaches $\sim$15 $Z_{\odot}$, or higher, and these quasars are classified as 
``N-loud quasars'' (e.g., \citealt{2003ApJ...583..649B,2004AJ....127..576B,
2004AJ....128..561B}). Here the definition of a N-loud quasar is that of
Jiang et al. (2008); i.e., type 1 quasars with strong nitrogen emission of 
$EW_{\rm rest}$(N~{\sc iv}]$\lambda$1486) $>$ 3 $\rm \AA$ or 
$EW_{\rm rest}$(N~{\sc iii}]$\lambda$1750) $>$ 3 $\rm \AA$ (see Section 2).
Note that the measurement of the N~{\sc v}$\lambda$1240 flux is sometimes 
uncertain due to the heavy blending with the Ly$\alpha$ emission 
\citep[see, e.g.,][]{2006A&A...447..157N}; however the N-loudness of these
quasars is convincing since the well-isolated N~{\sc iv}]$\lambda$1486
and N~{\sc iii}]$\lambda$1750 are used for defining the N-loud quasar population.
Standard chemical evolution models can not predict such extremely high 
metallicities at any epoch \citep[e.g.,][]{2008A&A...478..335B}, thus the 
presence of these N-loud quasars is a great challenge for galaxy evolutionary 
models.

\citet{2008ApJ...679..962J} pointed out that N-loud quasars may not have such  
high $Z_{\rm BLR}$, but simply have unusually high relative abundance of 
nitrogen in the BLR,  mainly because the emission-line spectrum of N-loud 
quasars is not significantly different from that of typical quasars except for the 
nitrogen lines. This idea poses into question the use of the N~{\sc v} emission 
for $Z_{\rm BLR}$ measurements in quasars. It is therefore very important to 
verify observationally if N-loud quasars have either high $Z_{\rm BLR}$ or just 
high nitrogen relative abundances. More recently, \citet{2011A&A...527A.100M} 
reported that the strength of nitrogen lines are correlated with the quasar 
Eddington ratio ($L_{\rm bol}/L_{\rm Edd}$) and discussed the possibility that 
quasars with strong nitrogen emission are in a special phase of the evolutionary 
history of their host galaxies. These studies outline the importance of 
understanding the nature of N-loud quasars, especially in terms of their chemical 
evolution. Therefore it is important to assess the metallicity of BLRs and host 
galaxies of N-loud quasars with a method different from that relying on rest-frame 
UV broad nitrogen lines.

In this paper, we focus on the  emission of the narrow-line region (NLR) in 
N-loud quasars. Since strong nitrogen lines from their BLRs suggest a high 
$Z_{\rm BLR}$, we examine the properties of the NLR to independently
investigate the metallicity in the N-loud quasar population. While the BLR is 
located in very compact, subpc-scale regions around SMBHs, NLR gas clouds 
are distributed on kpc scales, i.e. on the scales of the host galaxies, thus
enabling us to study the possible relationships between the properties of 
N-loud quasars and their host galaxies. Throughout this paper, we adopt a 
cosmology with $H_0 = 70$ km s$^{-1}$ Mpc$^{-1}$, $\Omega_m = 0.3$, and 
$\Omega_\Lambda = 0.7$.

\section{Sample Selection, Observation, and the Data Reduction}

\begin{table*}
\caption{Observational properties of SDSS J1707+6443}
\centering
\begin{tabular}{ccccccccc} 
  \hline\hline
  name & 
  $z$  &
  $i^{\prime}_{\mathrm{AB}}(\mathrm {mag})$ & 
  ${i^{\prime}}_{\mathrm {\scriptsize abs, AB}}(\mathrm {mag})^\mathrm{a}$ & 
  $H_{\mathrm{Vega}}(\mathrm {mag})^\mathrm{b}$ & 
  \multicolumn{4}{c}{$EW_{\rm rest}$ ($\AA$)$^\mathrm{c}$}
  \\                     
  & & & & &    
  N~{\sc iv}] & 
  N~{\sc iii}] & 
  C~{\sc iv} & 
  C~{\sc iii}]      
  \\ 
  \hline  
  SDSS J1707+6443 & 
  3.163 & 
  18.33 & 
  $-$28.06 & 
  16.15 & 
  5.1 & 
  2.5 & 
  40.8 & 
  15.3 
  \\ 
  \hline
\end{tabular}
\begin{list}{}{}
\item[$^\mathrm{a}$]
  $i^\prime$ band absolute magnitude given by \citet{2007AJ....134..102S}.
\item[$^\mathrm{b}$]
  2MASS magnitude given by \citet{2007AJ....134..102S}.
\item[$^\mathrm{c}$]
  Rest-frame equivalent widths of some BLR lines given by 
  \citet{2008ApJ...679..962J}.
\end{list}
\label{tb:proper}
\end{table*}

\begin{figure}
\includegraphics[width=9cm]{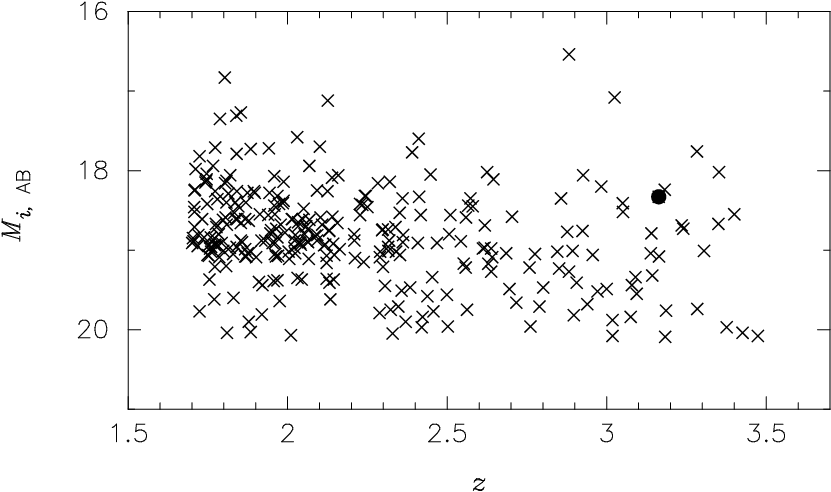}
\caption{
Absolute $i^\prime$-band magnitudes of 293 N-loud quasars, as a 
function of redshift. Filled circle and crosses denote SDSS J1707+6443
and the other N-loud quasars, respectively. Data are taken from
\citet{2008ApJ...679..962J}.
}
\label{fig:zMi}
\end{figure}

\begin{figure}
\includegraphics[width=9cm]{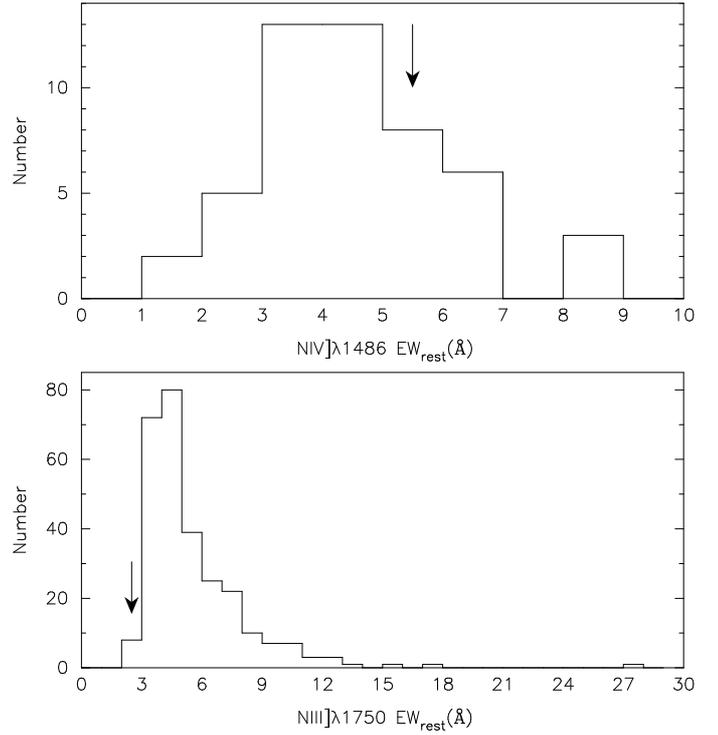}
\caption{
The histograms of $EW_{\rm rest}$ of N~{\sc iv}]$\lambda$1486 
(upper panel) and N~{\sc iii}]$\lambda$1750 (lower panel). The arrows 
denote the $EW~{\rm rest}$ values of SDSS J1707+6443. 
Data are taken from \citet{2008ApJ...679..962J}.
}
\label{fig:EW_N}
\end{figure}

\begin{figure}
\includegraphics[width=9cm]{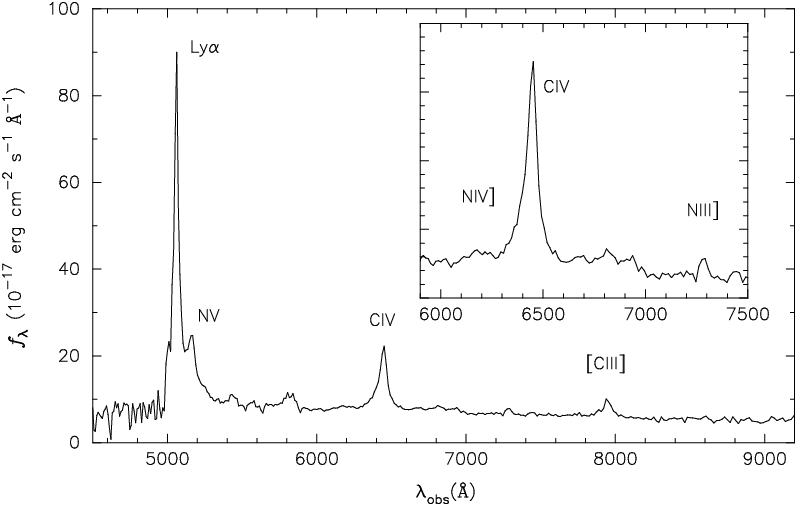}
\caption{
  Optical spectrum of SDSS J1707+6443, obtained from the SDSS database.
  Line IDs are given for strong emission lines.
  The inset shows spectral features around the C~{\sc iv} emission, where
  N~{\sc iv}]$\lambda$1486 and N~{\sc iii}]$\lambda$1750 lines are seen. 
}
\label{fig:dual}
\end{figure}

In order to investigate NLR properties of N-loud quasars in detail, we focus 
on SDSS J1707+6443 at $z = 3.163$ from the N-loud quasar catalog of 
\citet{2008ApJ...679..962J}. This catalog contains 293 N-loud quasars 
selected from the Fifth Data Release quasar catalog 
\citep{2007AJ....134..102S} of the Sloan Digital Sky Survey (SDSS; 
\citealt{2000AJ....120.1579Y}). The N-loud quasars in the catalog of 
\citet{2008ApJ...679..962J} were selected with the following criteria; (i) 
$i^\prime < 20.1$, (ii) $1.7 < z < 4.0$, and (iii) 
$EW_{\rm rest}$(N~{\sc iv}]$\lambda$1486) $>$ 3 ${\rm \AA}$ or
$EW_{\rm rest}$(N~{\sc iii}]$\lambda$1750) $>$ 3 ${\rm \AA}$. Among the 
N-loud quasars listed in the catalog of \citet{2008ApJ...679..962J}, 
we focused on SDSS J1707+6443 because it is relatively bright and 
its redshift is adequate for detecting redshifted NLR lines without suffering 
from the deep atmospheric absorption (Figure \ref{fig:zMi}). In Figure 
\ref{fig:EW_N}, we show the frequency distributions of 
$EW_{\rm rest}$(N~{\sc iv}]$\lambda$1486) and 
$EW_{\rm rest}$(N~{\sc iii}]$\lambda$1750), and also show where SDSS 
J1707+6443 is located in these frequency distributions (see also Table 1). 
The averages and standard deviations of 
$EW_{\rm rest}$(N~{\sc iv}]$\lambda$1486) distributions are 4.6$\rm \AA$ 
$\pm$ 1.7$\rm \AA$, with minimum and maximum values in the parent 
sample of 1.2$\rm \AA$ and 8.9$\rm \AA$. The averages and standard 
deviations of $EW_{\rm rest}$(N~{\sc iii}]$\lambda$1750) are 5.6$\rm \AA$ 
$\pm$ 2.6$\rm \AA$ , with minimum and maximum values of
2.5$\rm \AA$ and 27.1$\rm \AA$. Figure 2 does not suggest that SDSS 
J1707+6443 is located at the lowest end in the distribution of 
$EW_{\rm rest}$(N~{\sc iii}]$\lambda$1750), because some of N-loud 
quasars do not always show detectable emission lines of both 
N~{\sc iv}]$\lambda$1486 and N~{\sc iii}]$\lambda$1750. 
More specifically, only 50 N-loud quasars out of 293 show detectable 
N~{\sc iv}]$\lambda$1486 emission while 280 out of 293 show detectable 
N~{\sc iii}]$\lambda$1750 emission \citep{2008ApJ...679..962J}.
Although SDSS J1707+6443 has a high 
N~{\sc iv}]$\lambda$1486/N~{\sc iii}]$\lambda$1750 ratio
with respect to typical N-loud quasars, we do not discuss this issue in this 
paper since it is beyond the scope of this work. The SDSS spectrum and 
the basic observational properties of SDSS J1707+6443 are shown in 
Figure \ref{fig:dual} and Table \ref{tb:proper}.

To assess the NLR properties of type 1 quasars, it is necessary to 
investigate rest-frame optical spectra and thus near-infrared spectroscopic 
data are required for this target. Therefore we observed SDSS J1707+6443 
with MOIRCS \citep{2006SPIE.6269E..38I,2008PASJ...60.1347S}, the 
near-infrared spectrograph at the Subaru Telescope, on 30th May 2009. 
We used the HK grism, that results into the wavelength coverage 
$1.4\mu{\rm m} \lesssim \lambda_{\rm obs} \lesssim 2.3\mu{\rm m}$. By 
using the 0.6$^{\prime\prime}$-wide slit, the resulting spectral resolution is 
$R \sim 570$, which is measured through the widths of the observed OH 
airglow emission. The total integration time was 1800 sec, consisting of six 
300 sec independent exposures. We also observed HIP 86687 (A2 star with 
$H_{\rm Vega} = 8.9$ and an assumed effective temperature of 8810 K) for 
the flux calibration and the telluric absorption correction. During the 
observation, the typical seeing size was 0.8$^{\prime\prime}$ in the optical.

Standard data processing was performed using the available 
IRAF\footnote{
  IRAF is distributed by the National Optical Astronomy Observatory, 
  which is operated by the Association of Universities for Research 
  in Astronomy (AURA) under cooperative agreement with the National 
  Science Foundation.
} routines. Frames were flat-fielded using domeflat images, and sky 
emission was performed by subtracting pairs of subsequent frames 
with the target at different positions along the slit. Cosmic-ray events 
were then removed by using the \texttt{lineclean} task. We extracted 
the one-dimensional spectrum of the target using the \texttt{apall} task, 
with an aperture size of $\pm 5$ pixels (1.17$^{\prime\prime}$) from the
emission center and, in this process, sky residuals were removed.
Wavelength calibration was performed using OH sky lines. Finally, flux 
calibration and telluric absorption correction were carried out using the 
observed spectra of HIP 86687. The final processed spectrum was 
obtained by combining all single exposure frames.

To check the consistency of the flux calibration between the SDSS data 
and our MOIRCS data, we extrapolated the UV continuum emission 
toward the longer wavelength and compared the flux density at 
$\lambda_{\rm rest}$ = 5100$\rm \AA$ between the extrapolated spectrum 
and our MOIRCS spectrum. The adopted spectral index is derived by 
fitting the SDSS spectrum at the wavelength regions where strong 
emission-line features are not present ($\lambda_{\rm rest} \sim 
1350{\rm \AA}, 1450{\rm \AA}, 1670{\rm \AA}$, and 1970${\rm \AA}$). Here 
we assume a constant spectral index at 
$\lambda_{\rm rest} < 4000 {\rm \AA}$ \citep[see][]{2001AJ....122..549V}. 
The discrepancy in the continuum flux is $\sim$10\%, and therefore we
conclude that the flux calibration is consistent between the SDSS spectrum 
and our MOIRCS spectrum.
    
\section{Result}

\begin{figure}
\includegraphics[width=9cm]{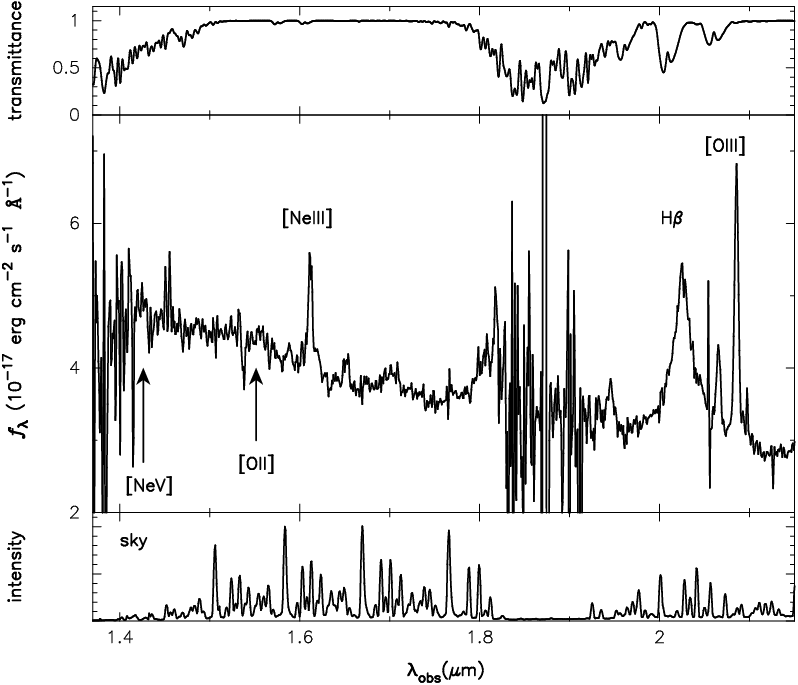}
\caption{
  Near-infrared spectrum of SDSS J1707+6443 obtained in our MOIRCS run
  (middle panel), with the Mauna Kea atmospheric transmission (upper panel)
  and the typical sky spectrum obtained during our MOIRCS run (lower panel).
  Important emission lines are labeled in the middle panel.
}
\label{fig:sdss}
\end{figure}

The final processed spectrum of SDSS J1707+6443 is shown in Figure 
\ref{fig:sdss}, with the Mauna Kea atmospheric transmission\footnote{
     Data obtained from the UKIRT web site.
} and a typical sky spectrum obtained during our MOIRCS observing run.
A prominent broad H$\beta$ emission is detected at $\lambda_{\rm obs} 
\sim$ 2.0 $\mu$m, a typical spectral feature from quasar BLRs. In addition to 
the broad H$\beta$ emission, some narrow emission lines from the NLR are 
also detected like the [O~{\sc iii}] doublet in the $K$-band, a feature common 
in several high-$z$ quasars \citep[e.g.,][]{1999ApJ...514...40M,
2004ApJ...614..558N,2009A&A...495...83M,2010ApJ...709..937G}. 
Interestingly, we also detect the quite rare [Ne~{\sc iii}]$\lambda$3869 
emission line but not the [Ne~{\sc v}]$\lambda$3426 and 
[O~{\sc ii}]$\lambda$3727 lines. 

\begin{figure}
\includegraphics[width=9cm]{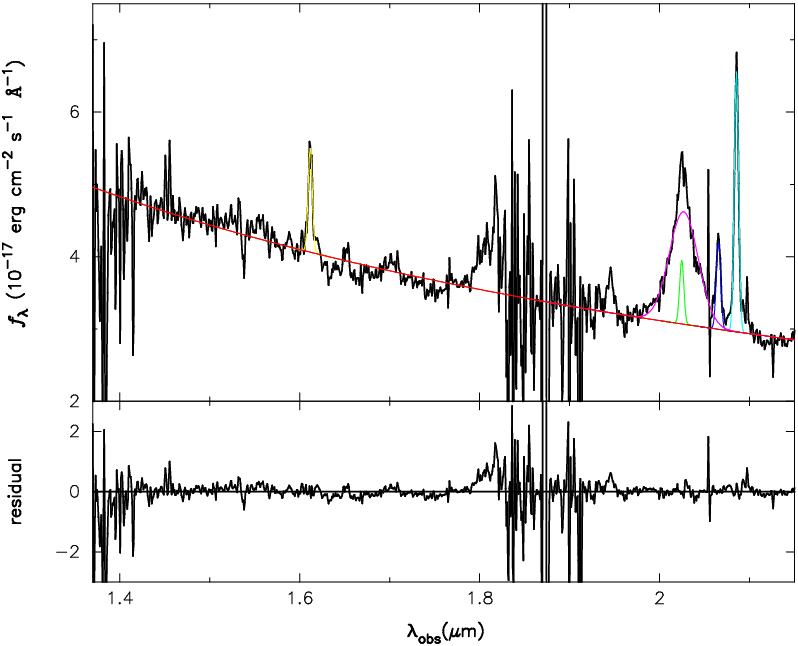}
\caption{
The fitting result (upper panel) and the residual (lower panel)
for the processed MOIRCS spectrum. See the main text for
details of the fitting procedure.
}
\label{fig:fit}
\end{figure}

To measure the emission-line fluxes and velocity widths, we fit the spectral 
features by using the \texttt{specfit} routine \citep{1994adass...3..437K}. Here 
we adopt a single Gaussian profile for forbidden narrow lines and a double 
Gaussian profile for H$\beta$ line. The measured emission-line properties 
are summarized in Table \ref{tb:emission}, where the presented 
quantities are based on the fitting models, not based on the actual data 
themselves. In Figure \ref{fig:fit}, we show fitting results and residuals.
The measured velocity widths ($\Delta v_{\rm FWHM}$ $\sim$ 600-850 
km s$^{-1}$ for the NLR lines and $\Delta v_{\rm FWHM}$ $\sim$ 5500 
km s$^{-1}$ for the BLR line) are consistent with typical 
values for NLR and BLR emission lines seen in type 1 AGNs 
(426 $\pm$ 251 km s$^{-1}$ and 4420 $\pm$ 3210 km s$^{-1}$ for the NLR 
and BLR lines; \citealt{2009ApJS..184..398H}). 
Note that the systematic errors given for the emission-line widths are 
estimated by applying some fitting functions for emission lines (such as 
Gaussian, Lorentzian, and so on) and examining the standard deviation 
of the width. Here the statistical errors in the velocity width are smaller
than the systematic errors, given the achieved signal-to-noise ratio.
   
\begin{table*}
\caption{The measured spectral features
}
\centering
\begin{tabular}{lccccc} 
\hline\hline
  line ID & center wavelength & EW$_{\rm rest}$ & line flux & FWHM$^\mathrm{a}$ & FWHM$^\mathrm{b}$ \\
  & [\AA] & [\AA] & [10$^{-17}$ erg s$^{-1}$ cm$^{-2}$] & [\AA] & [km s$^{-1}$] \\
\hline
  {}[Ne~{\sc v}]$\lambda$3426	& ---     & --- &$<$12.3$^{\rm c}$&---&---\\
  {}[O~{\sc ii}]$\lambda$3727	& ---     & --- &$<$ 10.0$^{\rm c}$&---&---\\
  {}[Ne~{\sc iii}]$\lambda$3869	& 16117.5 & 4.9$\pm$0.2 & 83.3$\pm$3.7 & 54.5$\pm$6.5 & 866 $\pm$103\\
  {}H$\beta_{\rm narrow}$		& 20245.7 & 4.1$\pm$0.2 & 51.8$\pm$2.3 & 55.3$\pm$8.4 & 627$\pm$95 \\
  {}H$\beta_{\rm broad}$		& 20267.1 &49.3$\pm$0.5 &628.6$\pm$5.9 &379.9$\pm$67.8 & 5595$\pm$999 \\
  {}[O~{\sc iii}]$\lambda$4959   & 20653.8  &5.8$\pm$0.1   &72.5$\pm$1.7    &56.4$\pm$8.6      &627$\pm$96\\
  {}[O~{\sc iii}]$\lambda$5007	& 20853.7 &17.7$\pm$0.1& 217.9$\pm$1.7 & 57.0$\pm$8.7 & 627$\pm$96 \\
\hline  
\end{tabular}
\begin{list}{}{}
\item[$^\mathrm{a}$]
  Measured emission-line width before the correction for the
  instrumental broadening.
\item[$^\mathrm{b}$]
  Emission-line velocity width after the correction for the instrumental broadening.
\item[$^\mathrm{c}$]
  For the undetected lines, $3 \sigma$ upper-limit fluxes are given.
\end{list}
\label{tb:emission}
\end{table*}

\section{Discussion}

\subsection{The Chemical Properties of the Narrow Line Region}

One of the aims of this study is to estimate the metallicity of the host galaxy of 
a N-loud quasar, SDSS J1707+6443. More specifically, we wish to test 
whether this quasar has extremely metal-rich gas clouds, as expected by its
strong broad nitrogen emission lines. Although strong broad nitrogen 
lines are usually interpreted as indications of high $Z_{\rm BLR}$ (for both 
permitted lines and semi-forbidden lines; see, e.g., 
\citealt{1976ApJ...204..330S,1992ApJ...391L..53H}), \citet{2008ApJ...679..962J} 
pointed out that  N-loud quasars might simply have unusually high nitrogen 
relative abundances. This is because the emission-line spectrum of the N-loud 
quasars is similar to that of usual type 1 quasars. In particular, many  N-loud 
quasars do not show anomalous behavior for what concerns other broad 
emission-line flux ratios, not involving nitrogen, but sensitive to $Z_{\rm BLR}$, 
such as (Si~{\sc iv}+O~{\sc iv}])/C~{\sc iv} \citep{2006A&A...447..157N,
2009A&A...494L..25J,2010MNRAS.407.1826S}. However, the nuclear BLR 
involves only a very small fraction of the gas content in the galaxy and, 
therefore, may well not be representative of the metallicity in the host galaxy. 
As a consequence, it may be more instructive to investigate the metallicity in 
the NLR ($Z_{\rm NLR}$) through the rest-frame optical spectrum, so to 
investigate whether the high metallicities inferred for the BLR are confirmed on 
the larger scales traced by the NLR.

\begin{figure}
\includegraphics[width=9cm]{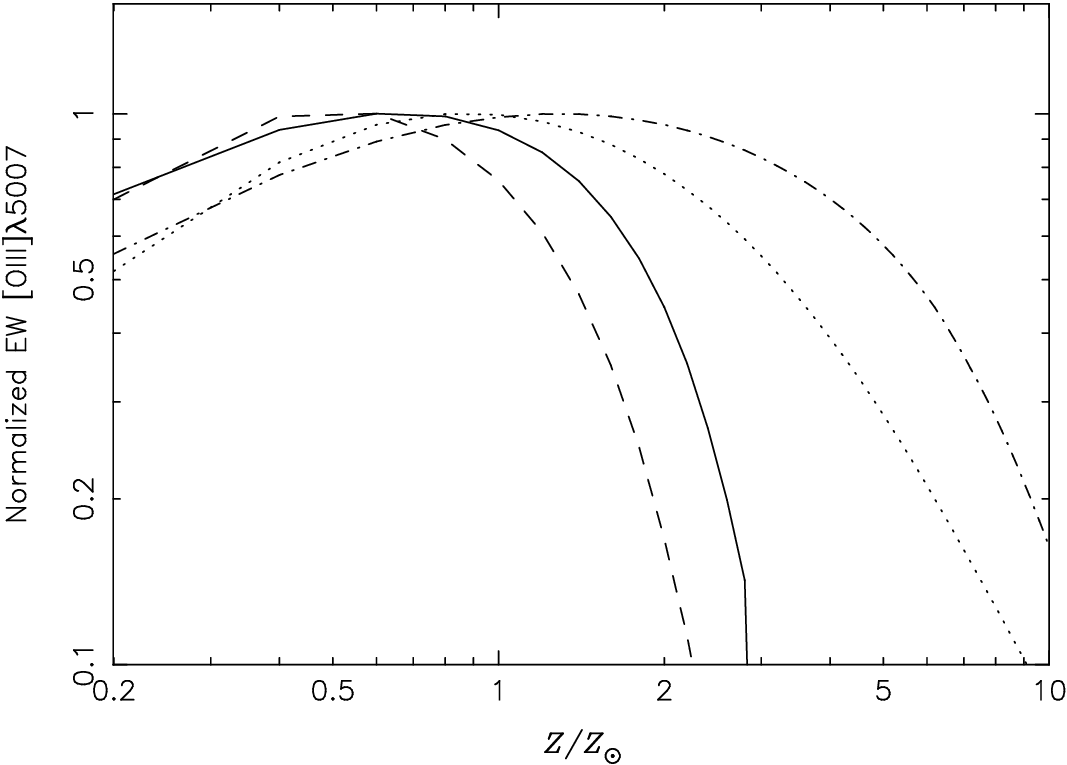}
\caption{
  Predicted equivalent width of [O~{\sc iii}]$\lambda$5007,
  as a function of $Z_{\rm NLR}$. Solid, dashed, dot-dashed, and
  dot lines denote the models with ($\log n_{\rm H}$, $\log U$) =
  (1, --1.5), (1, --3.5), (4, --1.5), and (4, --3.5), respectively.
  The EW predictions are normalized by their peak values.
}
\label{fig:metalmetal}
\end{figure}

We focus first on the metallicity dependence of line emissivity on the physical 
conditions typical of the NLR. The emissivity of collisionally excited emission 
lines strongly depends on the gas temperature. The equilibrium temperature 
of ionized gas clouds depends on gas metallicity, because metal emission 
lines are the main coolants of these clouds. Therefore, collisionally excited 
emission lines from the NLR become fainter for increasing $Z_{\rm NLR}$ 
\citep[e.g.,][]{2006A&A...447..863N,2009A&A...503..721M}. To show this effect 
more explicitly, we computed simple photoionization model calculations by 
using Cloudy version 08.00 \citep{1998PASP..110..761F,2006hbic.book.....F}. 
We assume ionization-bounded plane-parallel gas clouds with a constant 
density, photoionized by the typical spectral energy distribution for quasars
\citep{1987ApJ...323..456M,2006hbic.book.....F}. We examine the metallicity 
dependence of the equivalent width (EW) of the [O~{\sc iii}] emission, with the 
relative elemental abundance ratios fixed to the solar values except for 
nitrogen. We assume that the nitrogen relative abundance (N/H) scales with 
the square of the metallicity. 
In Figure \ref{fig:metalmetal}, we show photoionization models results
for gas densities of $n_{\rm H}$ = 10$^1$ cm$^{-3}$ and 10$^4$ cm$^{-3}$, 
and ionization parameters of $U = 10^{-3.5}$ and $10^{-1.5}$. The emissivity 
curves in this figure are normalized by their peak to highlight their dependence 
on the gas metallicity. Clearly the $EW$([O~{\sc iii}]) decreases rapidly as the 
metallicity increases, and the resultant $EW$([O~{\sc iii}]) is quite small at 
$Z_{\rm NLR}$ $>$ 5 $Z_{\odot}$. Note that we cannot derive accurate values 
of $Z_{\rm NLR}$ based only on [O~{\sc iii}], since detailed metallicity studies 
requires multiple emission lines \citep[e.g.,][]{1998AJ....115..909S,
2002ApJ...575..721N,2003ApJ...598..178I,2006MNRAS.371.1559G,
2006A&A...447..863N,2009A&A...503..721M}. However, the aim of this paper 
is only to assess whether the NLR in SDSS J1707+6443 has a high metallicity,
not to derive its accurate value. These results suggest that we can discriminate 
whether or not N-loud quasars have a relatively high $Z_{\rm NLR}$ with 
respect to other type 1 quasars by comparing the $EW$([O~{\sc iii}]) 
distributions of the two populations. 

\begin{figure}
\includegraphics[width=9cm]{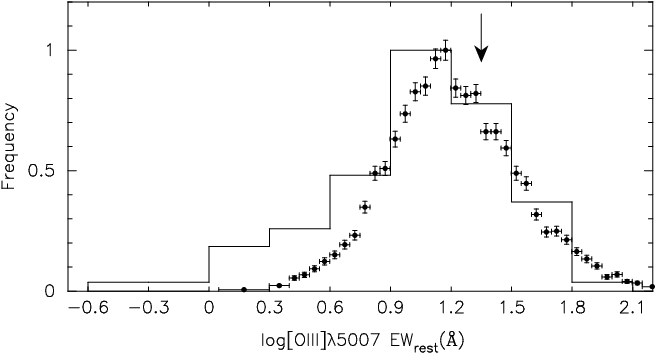}
\caption{
   Solid-line histogram denotes the compiled data of high-$z$ type 1 quasars 
   from the literature \citep{1999ApJ...514...40M,2004ApJ...614..558N,
   2009A&A...495...83M,2010ApJ...709..937G} Filled circles with error bars 
   denote the data of SDSS type 1 quasars at $0 < z < 1$
   \citep{2011MNRAS.411.2223R}. The frequency distributions of the two 
   datasets are normalized by the number of objects in the bin at the peak of 
   their frequency distributions. The arrow denotes the 
   EW$_{\rm rest}$([O~{\sc iii}]) value of SDSS J1707+6443 measured in our 
   work.
}
\label{fig:hist}
\end{figure}

In Figure \ref{fig:hist}, we show the frequency distribution of 
$EW_{\rm rest}$([OIII]) of type 1 quasars at 1 $<$ $z$ $<$ 4, that are compiled
from the literature (\citealt{1999ApJ...514...40M,2004ApJ...614..558N,
2009A&A...495...83M,2010ApJ...709..937G}; shown with a solid histogram in 
Figure \ref{fig:hist}). In these four papers, there are 86 [O~{\sc iii}]-detected 
quasars, whose average and median EW([O~{\sc iii}])$_{\rm rest}$ values are 
16.5 $\rm \AA$ and 13.0 $\rm \AA$ with the standard deviation of 14.6$\rm \AA$ 
(the minimum and maximum values are 0.4$\rm \AA$ and 73.0$\rm \AA$, 
respectively). Since the $EW_{\rm rest}$([O~{\sc iii}]) of SDSS J1707+6443 is 
17.4${\rm \AA}$ (Table \ref{tb:emission}), the [O~{\sc iii}] emission of SDSS 
J1707+6443 is not weaker than that of typical type 1 quasars. In Figure 
\ref{fig:hist} we also show the frequency distribution of 
$EW_{\rm rest}$([O~{\sc iii}]) of SDSS quasars at $0 < z < 1$ (shown with filled
circles with error bars in Figure \ref{fig:hist}). This is the distribution by 
\citet{2011MNRAS.411.2223R} who used the spectral measurements of SDSS 
DR5 quasars \citep{2007AJ....134..102S} performed by 
\citet{2008ApJ...680..169S}. The frequency distribution of 
$EW_{\rm rest}$([O~{\sc iii}]) of the $0 < z < 1$ SDSS sample
(\citealt{2011MNRAS.411.2223R}; filled circles in Figure \ref{fig:hist}) is not 
significantly different from that of the higher-$z$ ($1 < z < 4$) sample (solid 
histogram in Figure \ref{fig:hist}). The $EW_{\rm rest}$([O~{\sc iii}]) value of 
SDSS J1707+6443 is above the peak value of these distributions. Therefore 
we conclude that the EW$_{\rm rest}$([O~{\sc iii}]) of SDSS J1707+6443 is not 
significantly smaller than typical type 1 quasars in both low-$z$ and high-$z$ 
samples. Note that the actual relative [O~{\sc iii}]$\lambda$5007 strength of 
SDSS J1707+6443 with respect to the whole parent sample of type 1 quasars 
could be higher, because only [O~{\sc iii}]-detected quasars (i.e., relatively
strong [O~{\sc iii}] emitters) are selectively shown in Figure \ref{fig:hist}.

The relatively large $EW$([O~{\sc iii}])$_{\rm rest}$ of SDSS J1707+6443 with 
respect to the global population of AGNs (Figure \ref{fig:hist}) suggests that the 
NLR in SDSS J1707+6443 is not characterized by a very high $Z_{\rm NLR}$, 
when the photoionization model results shown in Figure \ref{fig:metalmetal} 
are taken into account. If a positive correlation between $Z_{\rm NLR}$ and 
$Z_{\rm BLR}$ is assumed, a lack of very high metallicity clouds in the NLR 
seems to be inconsistent with the strong broad nitrogen lines seen in the
rest-frame UV spectrum of this object. Obviously these are two possible 
scenario to explain this inconsistency: (1) the strong UV broad nitrogen lines 
are due to a very high relative abundance of nitrogen in the BLR (with respect 
to non-N-loud quasars) rather than to an extremely higher BLR metallicity than 
non-N-loud quasars, or (2) the BLR metallicity is significantly higher than the 
NLR metallicity, which better represents the metallicity of the host galaxy. The 
relation between $Z_{\rm BLR}$ and $Z_{\rm NLR}$ is not yet well understood
observationally, although \citet{2007ApJ...664L..75F,2008ApJ...677...79F,
2009ApJ...696.1693F} reported that these two quantities should be related with 
each other, at least in low-z ($z<0.5$) quasars. Whatever case (1) or (2) applies,  
the strong broad nitrogen lines of SDSS J1707+6443 are not consistent with the 
possibility that the host galaxy of this quasar is characterized by a very high 
metallicity. This implies that broad UV nitrogen lines of quasars are not (at least 
in some cases) a good tool to explore the chemical evolution of quasar host 
galaxies. Here we note that the spatial scale of NLRs is far larger than that of 
BLRs but not necessarily coincide with the spatial scale of their host galaxies; 
we show, however in the case of SDSS J1707+6443, the NLR spatial scale 
corresponds to the host-galaxy scale ($\sim$kpc scale; see Section 4.3). Since 
quasar spectra are frequently used to investigate the chemical evolution at
high redshifts, it will be crucial to examine whether the broad UV lines of 
quasars are good (or bad) tracers of the metallicity through more detail studies 
for larger samples of AGNs (see, e.g., \citealt{2011A&A...527A.100M}).

\subsection{Black Hole Mass and the Eddington Ratio}

The results in the previous section suggest that the strong nitrogen emission in 
the BLR of SDSS J1707+6443 (or, possibly, N-loud quasars in general) is not 
indicative of high metallicity in the NLR and hence in the host galaxy. It is 
however important to verify whether the properties of the broad nitrogen lines 
of SDSS J1707+6443 do follow the trends of the global population of high-z 
type 1 quasars. Recently \citet{2011A&A...527A.100M} found that a high relative 
nitrogen abundance is seen in BLRs of high-$L_{\rm bol}/L_{\rm Edd}$ quasars. 
To examine whether SDSS J1707+6443 is consistent with the result of 
Matsuoka et al. (2011), we estimate $M_{\rm BH}$ and $L_{\rm bol}/L_{\rm Edd}$ 
based on the velocity widths of C~{\sc iv} in the SDSS spectrum and H$\beta$ in
our MOIRCS spectrum. Note that it is debatable whether C~{\sc iv}-based or 
H$\beta$-based estimation are more accurate (see, e.g., 
\citealt{2004ApJ...614..558N,2004ApJ...613..682P,2009ApJ...702.1353D}). 
Therefore we use both emission lines to minimize possible systematic errors in 
the estimation of those parameters.

We derive $M_{\rm BH}$ for SDSS J1707+6443 by adopting the calibrations
given by \citet{2006ApJ...641..689V},
\begin{equation}
   \log \frac{M_{\rm BH}}{M_{\odot}} = 6.66 + 0.53 \log \left(
   \frac{\lambda L_\lambda (1350{\rm \AA})}{10^{44}{\rm \ erg \ s^{-1}}}
   \right) + 2 \log \left( \frac{FWHM}{{\rm 1000 \ km \ s}^{-1}} \right)
\end{equation}   
for C~{\sc iv}, and
\begin{equation}
   \log \frac{M_{\rm BH}}{M_{\odot}} = 6.91 + 0.50 \log \left(
   \frac{\lambda L_\lambda (5100{\rm \AA})}{10^{44}{\rm \ erg \ s^{-1}}}
   \right) + 2 \log \left( \frac{FWHM}{{\rm 1000 \ km \ s}^{-1}} \right)
\end{equation}
for H$\beta$, where $FWMH_{\rm CIV}$ and $FWHM_{H\beta}$ are the velocity 
width of the C~{\sc iv} and H$\beta$ emission in full-width at half maximum, 
respectively. For estimating the Eddington ratio ($L_{\rm bol}$$/$$L_{\rm Edd}$), 
we adopt a bolometric correction of $L_{\rm bol}$ = 9.26 
$\lambda L_\lambda$5100 (Shen et al. 2008). The derived values are 
log ($M_{\rm BH}$$/$$M_{\odot}$) = 8.98 and log ($L_{\rm bol}$$/$$L_{\rm Edd}$)
= $-$0.53 when using C~{\sc iv}, and log ($M_{\rm BH}/M_{\odot}$) = 9.73 and 
log ($L_{\rm bol}$$/$$L_{\rm Edd}$) = $-$0.21 when using H$\beta$, respectively, 
Accordingly we adopt their means, log ($M_{\rm BH}$$/$$M_{\odot}$) 
= 9.50 and log ($L_{\rm bol}$$/$$L_{\rm Edd}$) = $-$0.34. 
The derived Eddington ratio is relatively high (regardless of the adopted 
emission line for deriving $M_{\rm BH}$), with respect to the frequency 
distribution of the Eddington ratio of SDSS type 1 quasars at similar redshifts 
($-0.48 \pm 0.41$ for 3144 quasars at $3.0 < z < 3.3$; 
\citealt{2011ApJS..194...45S}).

Our results is consistent with the finding of \citet{2011A&A...527A.100M} that
quasars with high accretion rate are not necessarily characterized by high 
metallicities, but are characterized by high nitrogen abundances (resulting in 
stronger nitrogen lines). In the scenario proposed by \citet{2011A&A...527A.100M}, 
the high black hole accretion is delayed by a few 100 Myr, relative to the main 
episode of star formation, when intermediate-mass stars have evolved and have 
enriched the ISM with nitrogen.

\subsection{Emission-line Diagnostics in the Narrow Line Region}

By combining the emission-line fluxes of [O~{\sc iii}]$\lambda$5007 and 
[Ne~{\sc iii}]$\lambda$3869 and the upper limit on the flux of 
[O~{\sc ii}]$\lambda$3727, it is possible to investigate the properties of NLR gas 
clouds. This analysis enables us to study the physical properties of gas clouds 
in the host galaxy, because the NLR is extended on galactic scales in contrast to 
the BLR. Moreover, since the redshift of SDSS J1707+6443 ($z \sim 3$) 
corresponds to the peak of the global quasar activity  \citep[e.g.,][]
{2006AJ....131.2766R}, the gas properties of the quasar host galaxy are 
interesting to explore the interplay between AGNs and their host galaxies (i.e., 
the galaxy-SMBH coevolution). In this context, host galaxies of the N-loud 
quasars are particularly interesting since they may be in a special evolutionary 
stage, as mentioned in Section 4.2. Note that NLR emission lines other than 
[O~{\sc iii}]$\lambda$5007 have only rarely been observed in luminous high-z 
type 1 quasars, thus not allowing a detailed investigation of NLR gas properties
as done with our MOIRCS spectrum.

\begin{figure}
\includegraphics[width=9cm]{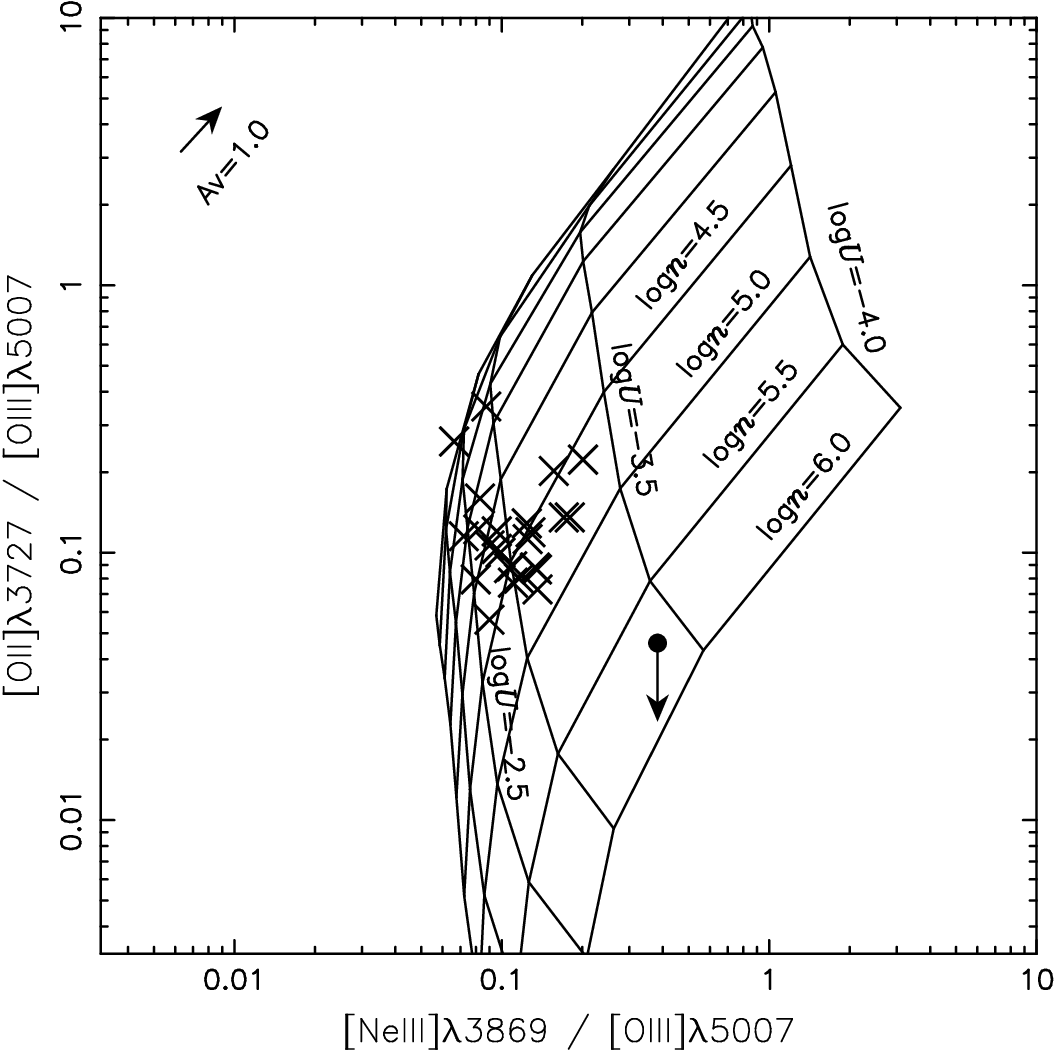}
\caption{
  Emission-line flux ratios of [O~{\sc ii}]$\lambda$3727/[O~{\sc iii}]$\lambda$5007 
  versus [Ne~{\sc iii}]$\lambda$3869/[O~{\sc iii}]$\lambda$5007. The filled circle 
  with an arrow denotes the data of SDSS J1707+6443 (where the 3$\sigma$ 
  upper limit is adopted for the [O~{\sc ii}] flux), and cross symbols denote the 
  data of the SDSS low-$z$ ($z \sim 0.7$) type 1 quasars. The arrow at the 
  upper-left corner in the panel is the reddening vector for observed data in the 
  case of $A_V = 1.0$ mag, adopting the extinction curve of 
  \citet{1989ApJ...345..245C}. The grids are predicted emission-line flux ratios 
  from Cloudy model runs, adopting $Z_{\rm NLR} = 3 Z_\odot$ and varying the 
  hydrogen density and the ionization parameter.
}
\label{fig:grid}
\end{figure}

In Figure \ref{fig:grid}, we place SDSS J1707+6443 on the diagram comparing
the distributions of the emission-line flux ratios of 
[O~{\sc ii}]$\lambda$3727$/$[O~{\sc iii}]$\lambda$5007 and 
[Ne~{\sc iii}]$\lambda$3869$/$[O~{\sc iii}]$\lambda$5007. For comparison, we 
also plot the data of the SDSS DR7 type 1 quasar sample (Shen et al. 2011). 
We checked their SDSS archival spectra and selected quasars whose 
[O~{\sc ii}]$\lambda$3727, [Ne~{\sc iii}]$\lambda$3869, and 
[O~{\sc iii}]$\lambda$5007 are significantly detected (i.e., S$/$N $>$ 10). As a
consequence we plot the emission-line flux ratio of 25 SDSS DR7 type 1 
quasars, whose redshift range is $0.37 < z < 0.80$ with the median redshift of 
0.65. Note that the average luminosity $\lambda$ $L_{\lambda}$(5100) for the 
25 SDSS type 1 quasars is 10$^{45.77}$ erg s$^{-1}$, that is $\sim$1 dex lower 
than the corresponding value for SDSS J1707+6443 (10$^{46.66}$ erg s$^{-1}$). 
As shown in Figure \ref{fig:grid}, these 25 SDSS type 1 quasars show completely 
different flux ratios from SDSS J1707+6443; i.e., SDSS J1707+6443 shows a 
higher [Ne~{\sc iii}]$\lambda$3869$/$[O~{\sc iii}]$\lambda$5007 ratio and a 
lower [O~{\sc ii}]$\lambda$3727$/$[O~{\sc iii}]$\lambda$5007 ratio than SDSS 
type 1 quasars. These differences can be interpreted qualitatively if the NLR in 
SDSS J1707+6443 is characterized by a much higher average density than the 
NLRs in SDSS type 1 quasars at $z < 1$. This is because the 
[O~{\sc ii}]$\lambda$3727 emission and partly [O~{\sc iii}]$\lambda$5007 
emission are suppressed by the collisional de-excitation process in high-density 
clouds with $n_{\rm H}$ $\sim$ 10$^6$ cm$^{-3}$ or higher, by taking the critical 
densities of these emission lines ($n_{\rm cr}$([O~{\sc ii}]) = (3.3-14) $\times$ 
10$^3$ cm$^{-3}$, $n_{\rm cr}$([O~{\sc iii}]) = 7.0 $\times$ 10$^5$ cm$^{-3}$, 
and $n_{\rm cr}$([Ne~{\sc iii]}) = 9.7 $\times$ 10$^6$ cm$^{-3}$) into account.

To investigate more quantitatively the differences in the NLR emission-line flux 
ratios between SDSS J1707+6443 and lower-$z$ type 1 quasars, we have 
performed Cloudy model runs for the parameter ranges of 
$n_{\rm H} = 10^2 - 10^6$ cm$^{\rm -3}$ and $U = 10^{-4.0} - 10^{-1.5}$. Here 
we adopt a metallicity of $Z_{\rm NLR} = 3 \ Z_\odot$, that is the typical value for 
NLR clouds \citep[e.g.,][]{2006MNRAS.371.1559G}, as we concluded in Section 
4.1 that SDSS J1707+6443 is not characterized by very high $Z_{\rm NLR}$. 
The other adopted parameters such as the input SED are the same as those 
described in Section 4.1. The results of the model runs are overlaid in Figure 
\ref{fig:grid}. The NLRs in the SDSS type 1 quasars are characterized by 
$n_{\rm H} \sim 10^{4.5}$ cm$^{-3}$ and $U \sim 10^{-3.5} - 10^{-2.5}$, 
consistent with the parameter ranges inferred by previous studies (e.g., 
\citealt{2001ApJ...546..744N}). On the other hand, the NLR in SDSS 
J1707+6443 is characterized by a higher density, $n_{\rm H}$ $\sim$ $10^6$ 
cm$^{-3}$ or more, although the ionization parameter is similar to that of type 1
SDSS quasars. These results are not sensitive to the adopted metallicity.

There are two possible scenarios to explain the distinct physical properties of 
SDSS J1707+6443. One possibility is considering a strong contribution from 
the inner wall of the dusty torus. Since the clouds located at the inner side of 
dusty tori are characterized by a high density and photoionized by the strong 
central continuum radiation, emission lines with high critical densities are 
radiated from these clouds (e.g., \citealt{1998ApJ...497L...9M,
2000AJ....119.2605N,2001PASJ...53..629N}). In the case that such high-density 
clouds at the inner dusty torus show strong NLR emission, its emission 
dominates the whole NLR emission and then low-$n_{\rm cr}$ emission lines 
become relatively weak, explaining the observed emission-line flux ratios of 
SDSS J1707+6443 (see also \citealt{1998ApJ...503L.115M,
2001ApJ...549..155N}). Another possibility is the existence of high-density 
clouds distributed on galactic scales (i.e., $\sim$kpc). Such a situation has 
been proposed for high-$z$ and/or high-luminosity quasars 
(\citealt{2004ApJ...614..558N}; see also \citealt{2005ApJ...629..680H}), possibly 
related to violent star-formation activity in the host galaxy. Note that the latter 
scenario is not ruled out by the non-detection of the [O~{\sc ii}]$\lambda$3727 
emission, which is used as an indicator of the star-formation rate 
\citep[e.g.,][]{1989AJ.....97..700G,1998ARA&A..36..189K,2009ApJ...700..971I}. 
This is because the [O~{\sc ii}]$\lambda$3727 emission is significantly 
suppressed in high-density H~{\sc ii} regions when the density is higher than 
$10^4$ cm$^{-3}$, even when the star-formation activity is very vigorous.

To distinguish between the above two scenarios, we estimated the typical 
distance of the ionized clouds from the nucleus in SDSS J1707+6443 by 
combining the estimates of $n_{\rm H}$ and $U$ obtained in the previous 
section. Since we already estimated $L_{\rm bol}$ of SDSS J1707+6443 in 
Section 4.1, the number of the ionizing photon can be estimated for a given 
spectral energy distribution (SED). Here we assume the same SED used in the 
above photoionization model calculations \citep{1987ApJ...323..456M} and 
adopt the parameters of  $n_{\rm H} = 10^6$ cm$^{-3}$ and $U = 10^{-3.5}$.
Accordingly, we obtain the result that the typical distance of the clouds is 
$\sim$1.5 kpc, which is consistent with the latter scenario that considers 
high-density clouds in the host galaxy. Note that a smaller radius would have 
been derived by assuming a higher density, e.g., $\sim$15 pc in the case of 
$n_{\rm H} \sim 10^8$ cm$^{-3}$. However this is not plausible, because 
[O~{\sc iii}]$\lambda$5007 emission would be significantly suppressed at such 
high-densities due to the collisional de-excitation. We thus conclude that there 
are high-density clouds at the kpc scale in the host galaxy of SDSS 
J1707+6443, that dominates the emission of the detected NLR lines. The 
torus scenario is disfavored also in terms of the observed NLR velocity width.
By assuming that the SMBH is the dominant source of the gravitational potential 
field in the nucleus of SDSS J1707+6443, the expected velocity width of 
[O~{\sc iii}]$\lambda$5007 is $\sim 1300$ km s$^{-1}$ if the most of this emission
arise at $\sim$15 pc from the SMBH with $M_{\rm BH} = 10^{9.78}M_{\odot}$
(see Section 4.2). This expected velocity width is apparently larger than the
observed width (617 km s$^{-1}$; Table \ref{tb:emission}), suggesting that 
most of the [O~{\sc iii}]$\lambda$5007 emission in this object arises at larger
spatial scales that probably corresponds to the scale of its host galaxy.

Here we briefly discuss the implication of the extended ($\sim$kpc) dense 
($\sim$10$^6$ cm$^{-3}$) gas clouds in SDSS J1707+6443 inferred from 
emission-line diagnostics. It is well known that such dense clouds in galaxies 
are closely related to the star-forming activity. 
\citet{1997ApJ...476..730P} reported their observations of the carbon
monosulphide (CS) molecule for $>$100 Galactic H~{\sc ii} regions and 
showed that their typical density is $n$ $\sim$ 10$^6$ cm$^{-3}$ (but less 
than 2 $\times$ 10$^7$ cm$^{-3}$). 
Such a high density for clouds in H~{\sc ii} 
regions has been inferred also from some other molecular-line radio 
observations (e.g., \citealt{1996ApJ...460..359H,1996ApJ...460..343B,
1997ApJ...488..286L}). Although such a high density is not the "typical" density 
for H~{\sc ii} region clouds, the star-formation efficiency of the denser clouds is 
actually much higher than less dense clouds. This is suggested by, e.g., a 
clear positive correlation between the $L_{\rm HCN}$$/$$L_{\rm CO}$ ratio 
(i.e., the dense-gas fraction) and the $L_{\rm IR}/L_{\rm CO}$ ratio (i.e., the 
star-formation efficiency) that is seen in H~{\sc ii} region clouds (e.g., 
\citealt{2004ApJ...606..271G}). Therefore high-density clouds generally 
dominate the star formation, when they exist (see also, e.g., 
\citealt{2011MNRAS.416L..21W}). A similar situation is also seen in high-z 
galaxies (e.g., \citealt{2007ApJ...660L..93G,2007ApJ...671L..13R,
2010ApJ...725.1032R,2011ApJ...726...50R}) but with a significant difference
compared to low-$z$ galaxies; that is the spatial extension of vigorous 
star-forming regions. The spatial scale of star-forming regions in high-$z$ 
actively star-forming galaxies (such as sub-millimeter galaxies) extends up to 
$\sim$kpc scales or more (e.g., \citealt{2004ApJ...611..732C,
2009Natur.457..699W}), that is different from low-$z$ actively star-forming 
galaxies such as ultra-luminous infrared galaxies (e.g., 
\citealt{1998ApJ...507..615D}; see also, e.g., \citealt{2009ApJ...695.1537I}). 
These pictures well match with our results in the sense that dense gas 
clouds ($\sim$10$^6$ cm$^{-3}$) are distributed at the $\sim$kpc scale in a 
galaxy at $z$ $\sim$ 3.2. Therefore, given these considerations, we speculate
the existence of vigorous star-forming activity in SDSS J1707+6443. 

\section{Summary}

To assess the physical and chemical properties of the host galaxies of N-loud 
quasars, we analyzed the MOIRC near-infrared spectrum of SDSS
J1707+6443, at $z$=3.16, obtaining the following results:

\begin{itemize}
\item This N-loud quasar shows strong [O~{\sc iii}]$\lambda$5007 emission. 
   Since photoionization models predict weak [O~{\sc iii}]$\lambda$5007 
   emission when the NLR metallicity is very high ($>$ 5 $Z_{\odot}$), the 
   detected strong [O~{\sc iii}]$\lambda$5007 emission suggests that the NLR in 
   this object is not characterized by very high metallicities.
\item The UV nitrogen lines from BLRs are not ideal tools to discuss the chemical 
   evolution of quasar host galaxies, because whatever the origin of strong broad 
   nitrogen emission (high $Z_{\rm BLR}$, or high relative abundance of nitrogen 
   with an ordinary metallicity), $Z_{\rm BLR}$ are likely unrelated with the 
   metallicities of the host galaxies.
\item The Eddington ratio of SDSS J1707+6443 is moderately high 
   ($L_{\rm bol}/L_{\rm Edd} = 10^{-0.26}$), as derived from single-epoch 
   $M_{\rm BH}$ estimates using both C~{\sc iv} and H$\beta$ emission lines. 
   This is consistent with the discovery by \citet{2011A&A...527A.100M} that a 
   high relative nitrogen abundance is associated to quasars with high Eddington 
   ratios.
\item The flux ratio of [O~{\sc ii}]$\lambda$3727/[O~{\sc iii}]$\lambda$5007 in 
   SDSS J1707+6443 is significantly lower than that in lower-$z$ type 1 quasars, 
   indicating that NLR clouds in SDSS J1707+6443 are characterized by much 
   higher densities than those in other type 1 quasars.
\item Photoionization models suggest that those high-density clouds (with 
   $n_{\rm H}$ $\sim$ 10$^6$ cm$^{-3}$) are located at the kpc scales in the host 
   galaxy of SDSS J1707+6443, and we speculate that this might be related to the 
   vigorous star-formation activity in the host galaxy.
\end{itemize}

These results possibly reveal a strong connection between the high AGN activity 
(characterized by a high Eddington ratio) and the vigorous star-formation activity 
suggested by the kpc-scale distribution of dense gas cloud in a N-loud quasar, 
SDSS J1707+6443. Currently it is not clear whether such a connection is seen 
also in other N-loud quasars generally or not, given a paucity of detailed 
near-infrared spectroscopic studies for N-loud quasars. Therefore it is highly 
interesting to examine whether such a situation is common in other N-loud 
quasars through further near-infrared spectroscopic observations for a large 
sample of N-loud quasars.

\begin{acknowledgements}
We thank the Subaru Telescope staff for supporting our MOIRCS observation.
We also thank the anonymous referee, Takayuki Saitoh, and Bunyo
Hatsukade, for their useful comments.
Funding for the creation and distribution of the SDSS Archive has been 
provided by the Alfred P. Sloan Foundation, the Participating Institutions, the 
National Aeronautics and Space Administration, the National Science 
Foundation, the U.S. Department of Energy, the Japanese Monbukagakusyo, 
and the Max Planck Society. The SDSS web site is http://www.sdss.org/. We 
thank Gary Ferland for providing his photoionization code Cloudy to the public.
N.A. is supported in part by a grant from the Hayakawa Satio Fund 
awarded by the Astronomical Society of Japan. T.N. is financially supported by
JSPS (grant no. 23654068), the Kurata Memorial Hitachi Science and 
Technology Foundation, the Itoh Science Foundation, Ehime University (the
Research Promotion Award), and Kyoto University (the Hakubi Project grant).
K.M. acknowledges financial support from JSPS through JSPS Research
Fellowships for Young Scientists. Y.T. is financially supported by JSPS 
(grant no. 23244031).
\end{acknowledgements}

\bibliographystyle{aa}

\end{document}